\pdfoutput=1
\documentclass[10pt,conference,compsocconf,letterpaper]{IEEEtran}
\usepackage{comment}
\usepackage{algorithmic}
\usepackage{hyperref}
\usepackage{amsmath}
\usepackage{xspace}
\usepackage{graphicx}
\usepackage{color}
\usepackage{caption}
\usepackage{subcaption}

\makeatletter
\newif\if@restonecol
\makeatother

\usepackage[ruled,vlined,nofillcomment]{algorithm2e}
\SetCommentSty{textit}
\SetKw{NEW}{new}
\SetKw{NOT}{not}
\SetKw{TO}{to}
\SetKw{OR}{or}
\SetKw{AND}{and}
\SetKwBlock{INIT}{Init:}{}
\SetKwBlock{LOCAL}{Local algorithm:}{}
\SetKwBlock{AGGR}{Aggregation algorithm:}{}

\SetKwFor{FUNCTION}{function}{}{}

\newcommand{\etsch}{\textsc{etsch}\xspace}
\newcommand{\dfep}{\textsc{dfep}\xspace}
\newcommand{\dfepc}{\textsc{dfepc}\xspace}

\newcommand{\Dist}{\textit{dist}\xspace}
\newcommand{\Source}{\textit{source}\xspace}
\newcommand{\Neighbors}{\textit{neighbors}\xspace}

\newcommand{\Id}{\textit{id}\xspace}

\newcommand{\owner}{\textit{owner}\xspace}
\newcommand{\eligible}{\textit{eligible}\xspace}
\newcommand{\funding}{\textit{funding}\xspace}

\newcommand{\Empty}{\textsf{\small empty}\xspace}
\newcommand{\Pop}{\textsf{\small pop}\xspace}
\newcommand{\Add}{\textsf{\small add}\xspace}
\newcommand{\Update}{\textsf{\small update}\xspace}
\newcommand{\MIN}{\textsf{\small min}\xspace}
\newcommand{\Init}{\textsf{\small init}\xspace}
\newcommand{\Local}{\textsf{\small localComputation}\xspace}
\newcommand{\Aggregation}{\textsf{\small aggregation}\xspace}
\newcommand{\Random}{\textsf{\small random}\xspace}

\newcommand{\Node}{\mathrm{Vertex}\xspace}
\newcommand{\ID}{\mathrm{ID}\xspace}
\newcommand{\Distance}{\mathrm{Distance}\xspace}
\newcommand{\PriorityQueue}{\mathrm{PriorityQueue}\xspace}

\newcommand{\dataset}[1]{\textsc{#1}}
\newcommand{\metric}[1]{\noindent\textbf{#1}:}

\begin{document}

\title{Distributed Edge Partitioning for Graph Processing}

\author{\IEEEauthorblockN{Alessio Guerrieri}
\IEEEauthorblockA{University of Trento\\
Email: a.guerrieri@unitn.it}
\and
\IEEEauthorblockN{Alberto Montresor}
\IEEEauthorblockA{University of Trento\\
Email: alberto.montresor@unitn.it}}

\maketitle

\begin{abstract}
The availability of larger and larger graph datasets, growing exponentially
over the years, has created several new algorithmic challenges to be addressed.
Sequential approaches have become unfeasible, while interest on parallel and distributed
algorithms has greatly increased.
Appropriately partitioning the graph as a preprocessing step can improve the degree of
parallelism of its analysis. A number of heuristic algorithms have been
developed to solve this problem, but many of them subdivide the graph on its
\emph{vertex} set, thus obtaining a vertex-partitioned graph.
Aim of this paper is to explore a completely different approach based
on \emph{edge partitioning}, in which edges, rather than vertices, are
partitioned into disjoint subsets. Contribution of this paper is twofold:
first, we introduce a graph processing framework based on
edge partitioning, that is flexible enough to be applied to several
different graph problems. Second, we show the feasibility of these
ideas by presenting a distributed edge partitioning algorithm called
\dfep.
Our framework is thoroughly evaluated, using both simulations and an Hadoop implementation running
on the Amazon EC2 cloud. The experiments show that \dfep is efficient, scalable
and obtains consistently good partitions. The resulting edge-partitioned graph
can be exploited to
obtain more efficient implementations of graph analysis algorithms.
\end{abstract}

\section{Introduction}
\label{sec:introduction}

One of the latest trend in computer science is the emergence of the ``big
data'' phenomena that concerns the retrieval, management and analysis of
datasets of extremely large dimension, coming from wildly different settings.
For example, astronomers need to examine the huge amount of observations collected by
the new telescopes that are being built both on Earth and in
orbit~\cite{hassan2011}. Biological experiments create large genomic and proteinomic
datasets that need to be processed and understood to reach new breakthroughs in
the study of drugs~\cite{howe2008}. Governments can improve the quality of life of
their citizens by analyzing the huge collections of individual events related
to traffic, economy, health-care and many other areas of everyday life~\cite{hardey2007}. The scale of such datasets
keeps increasing exponentially, moving from gigabytes to terabytes and now even to
petabytes. 

Although the collected data is often structured, several interesting datasets
are unstructured and can be modeled as graphs.
An obvious example is the World Wide Web, but
there are many other examples such as social network topologies, biological
systems or even road networks. While graph problems
have been studied since before the birth of computer science, the sheer size of these
datasets makes even classic graph problems extremely difficult. Even solving the
shortest path problem needs too many iterations to complete when the graph is too big
to fit into memory. The big Internet players (such as Google, Yahoo and
Facebook) have invested large amount of money in the development of novel
distributed frameworks for very large graph analysis and are working on new
solutions of many interesting classic problems in this new context~\cite{Malewicz2010,Bialecki2005}.

%
While parallel (multi-cpu, multi-core) systems have been used to deal with this
deluge of data, there are  many cases in which distributed approaches
are the only viable road. The disadvantages of distribution
cannot be ignored, though: they are inherently more difficult to develop
and implement, and they bring a larger communication overhead. Nevertheless, the advantages
outweigh the disadvantages. A distributed system is able
to cope with potentially unlimited datasets, is more robust to hardware failures, is often
cheaper and, with the emergence of distributed frameworks for data analysis,
is also much easier to use than it was a decade ago. These new distributed
frameworks abstract away most of the challenges of building a distributed
system and give to the analyst simple programming models to write
their data analysis programs.

A very common pattern in both parallel and distributed computing is to first
partition the data, then work on each partition separately, minimizing the
amount of communication between threads, processes or nodes. On graph datasets,
this typically means partitioning the vertices into non-overlapping
subsets, called \emph{partitions}. Edges between vertices that have
been assigned to distinct partitions act as communication channels between the
partitions themselves. 

When such partitions are assigned to a set of independent computing entities
(being them actual machines or virtual executors like processes and threads, or
even mappers and reducers in the MapReduce model), their size matters: the
largest of them must fit in the memory of a single computing entity. A common
solution to the problem of optimizing the usage of memory in such cases, is to compute
partitions that have similar sizes. Dividing the vertex set in equal sized partitions can still 
lead to an unbalanced subdivision, though: having the same amount of vertices does not imply that the
corresponding subgraph have the same size, given the unknown distribution of 
their edge degrees. 

In this paper we make the case for a different approach: edges are
partitioned into disjoint subsets, while vertices are associated to edges and
thus may belong to several partitions at the same time. Our contribution is
twofold: first, we propose a novel edge-based distributed graph processing
framework called \etsch, in which computation is associated to edges rather
than vertices, and we show that such framework can be used to compute the most
common properties of graphs. Second, we design the first building block of this
framework by proposing \dfep, a distributed edge-partitioning algorithm that
can be used in the pre-processing phase to obtain the disjoint edge partitions
to be fed into \etsch.

The paper thoroughly evaluates \etsch and \dfep, using both simulations
and an Hadoop implementation running on the Amazon EC2 cloud. The experiments
show that \dfep is efficient, scalable and obtains consistently good
partitions. The resulting edge-partitioned graph is easier to analyze than the
unpartitioned graph and can be exploited to obtain more efficient
implementations of graph analysis algorithms.

The rest of the paper is structured as follows. Section~\ref{sec:edgepart}
introduces the main concepts related to edge partitioning. Section~\ref{sec:framework}
shows how to organize a distributed computation over an edge-partitioned graph
in \etsch. We then move to our
second contribution by proposing,  in Section~\ref{sec:algorithm},  our novel distributed edge partitioning
algorithm called \dfep.
The experimental results are presented in Section~\ref{sec:results}.
Section~\ref{sec:stateart} presents the related work. The paper finishes with the
conclusions in Section~\ref{sec:conclusion}.

\section{Edge partitioning} \label{sec:edgepart}

The task of subdividing a graph into partitions of similar size, or \emph{partitioning}, is
a classical problem in graph processing, and has many clear applications in
both distributed and parallel graph algorithms. Most solutions, from Lin's and
Kernighan's algorithm~\cite{Kernighan1970} in the 70's to more recent
approaches~\cite{Tsourakakis2012}, try to solve \emph{vertex} partitioning. This
approach, however, may lead to unbalanced partitions, because even if they end
up in having the same amount of vertices, an unbalanced distribution of edges may
cause some subgraphs to be much larger than others. Approaching the problem
from an edge perspective, thus, may bring us to interesting and practical
results.

\begin{figure}
\begin{center}
\includegraphics[width=.43\textwidth]{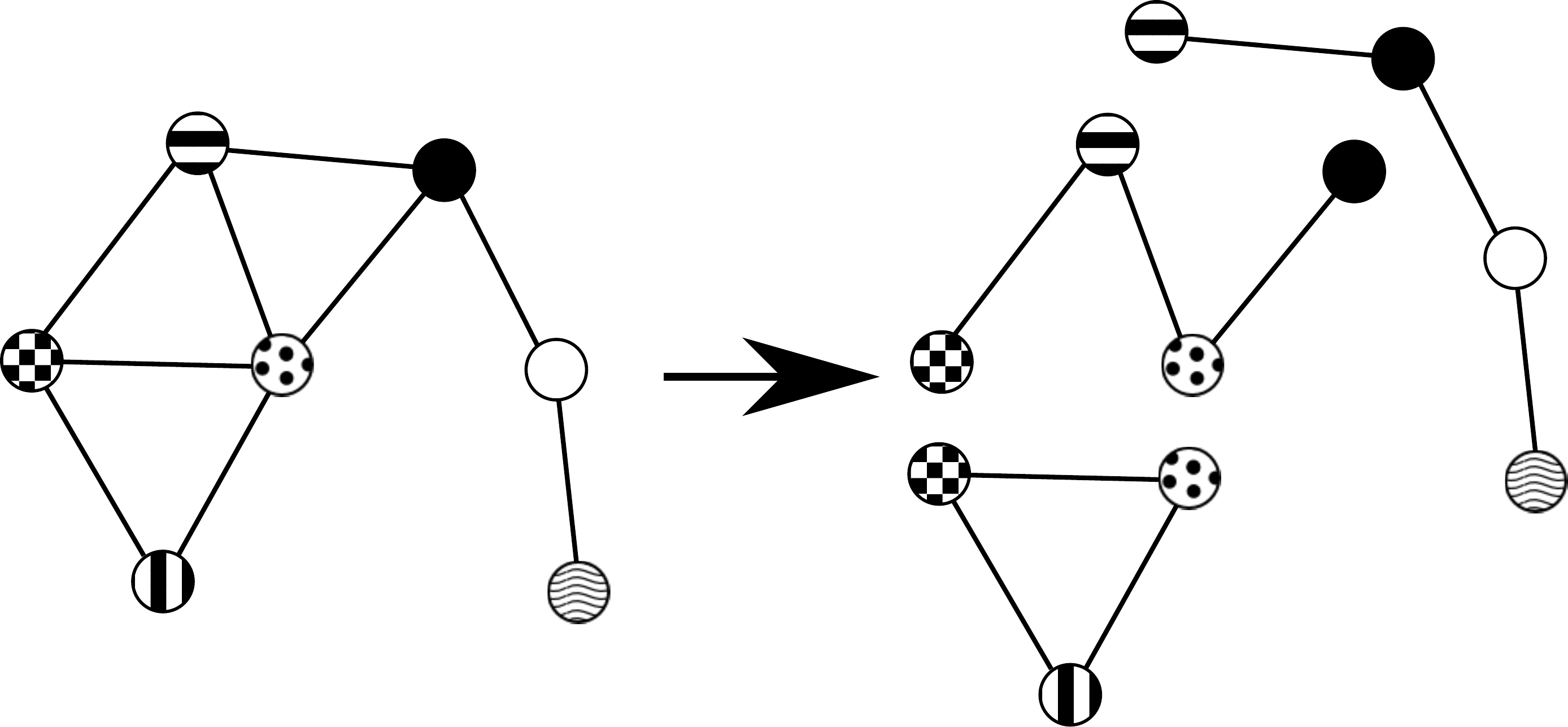}
\end{center}
\caption{Edge partitioning example: each edge appears in only one partition, while frontier vertices may appear in more than one partition}
\label{fig:example}
\end{figure}

Given a graph $G = (V,E)$ and a parameter $K$, an \emph{edge partitioning} of
$G$ subdivides all edges into a collection $E_1, \ldots, E_K$ of
non-overlapping edge partitions, i.e. 
\[
  E = \cup_{i=1}^K E_i \qquad 
  \forall i,j: i \neq j \Rightarrow E_i \cap E_j = \emptyset
\]
The $i$-th partition is associated with a vertex set $V_i$,
composed of the end points of its edges:
\[
   V_i = \{ u : (u,v) \in E_i \vee (v,u) \in E_i \}
\]
The edges of each partition, together with the associated vertices, form the
subgraph $G_i = (V_i, E_i)$ of $G$, as illustrated in Figure~\ref{fig:example}. 

The \emph{size} of a partition is proportional to the amount of edges and vertices
$|E_i|+|V_i|$ belonging to it. Given that each edge $(u,v) \in E_i$ contributes
with at most two vertices, $|V_i| = O(|E_i|)$ and the amount of memory needed
to store a partition is strictly proportional to the number of its edges. This
fact can be exploited to distribute fairly the load among machines.

Vertices may be replicated among several partitions, in which case are
called \emph{frontier vertices}. We denote with $F_i \subseteq V_i$ the set of vertices
that are frontier in the $i$th partition. These vertices are the channels through which
the partitions communicate.

\section{Working with edge partitionings}\label{sec:framework}

When a graph is subdivided using a vertex partitioning algorithm, each subgraph
has a number of external edges that connect vertices across partitions. These are
cut edges that are not really part of the subgraph, since the partition
does not have knowledge of the other endpoint of the edge. The approach in
this case is to consider vertices as computational entities that ``send'' messages
to their neighbors, potentially across partitions using the cut edges.

This is not the case with edge partitioning. Both vertices and edges of a local
graph can be associated with local state. Edges are part of exactly one
subgraph, so their state belongs exactly to one partition. The same happens
with vertices that do not belong to the frontier. Frontier vertices, on the
other hand, are replicated in different partitions and their state need to
be periodically reconciled. These idea are at the basis of \etsch, our graph
processing framework based on edge partitioning.

\begin{figure}[t]
\begin{center}
\includegraphics[width=.35\textwidth]{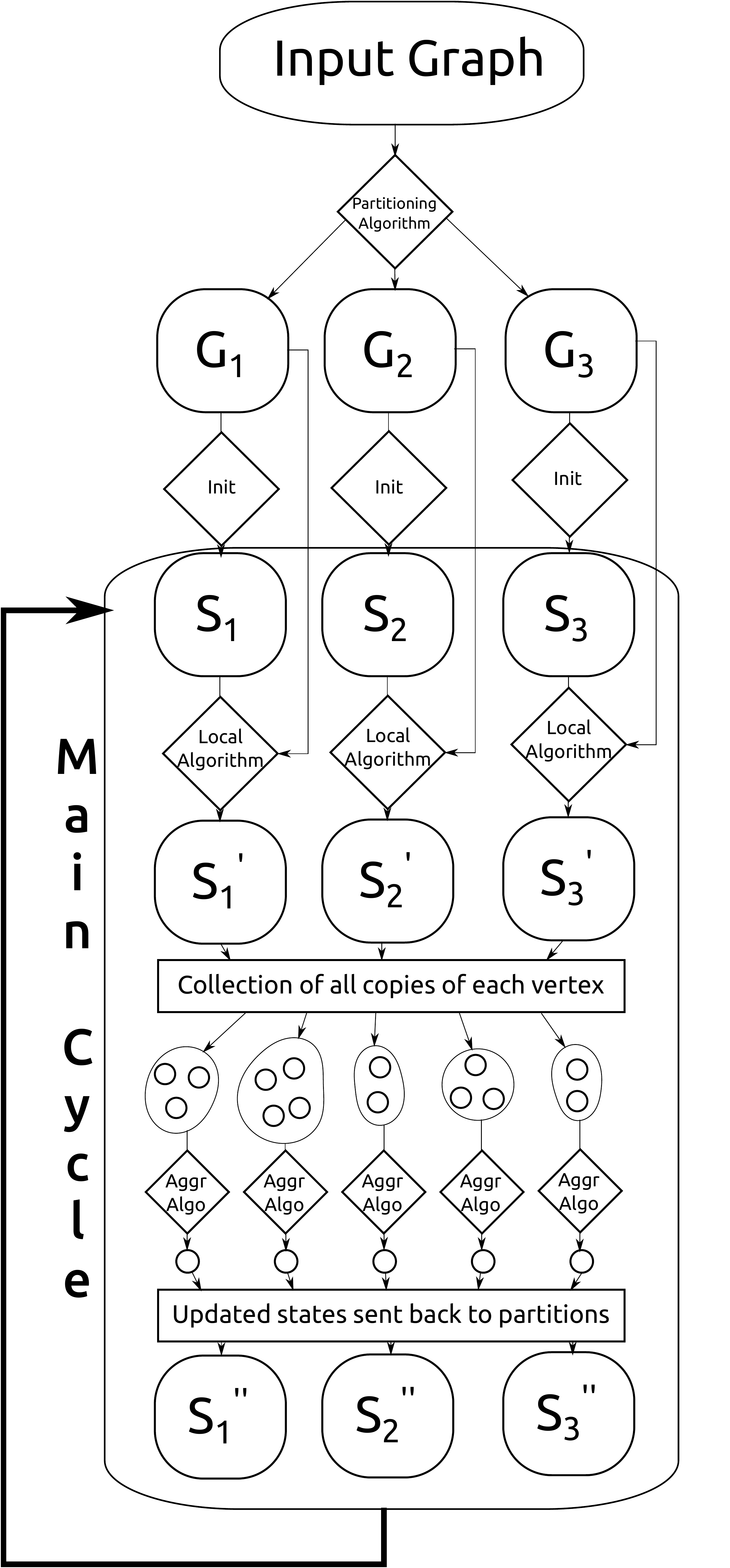}
\end{center}
\caption{Schema of our framework}
\label{fig:framework}
\end{figure}

Figure~\ref{fig:framework} shows the organization of \etsch. First of all, the graph
is decomposed into $K$ partitions by an edge partitioning algorithm like \dfep.
Each partition is assigned to a different
\emph{worker}, which executes the following steps:
\begin{enumerate}
\item The \emph{initialization phase} is run once, by taking the subgraph
representing the partition as input and initializing the local state of vertices
and edges.
\item Once completed the initialization, each subgraph state is fed to the 
\emph{local computation phase}, that runs an independent instance of a sequential 
algorithm that updates the local state of the subgraph.
\item The \emph{aggregation phase} logically follows the local computation: for
each frontier vertex, the framework collects the distinct states of all replicas
and computes a new state, that is then copied into the replicas. 
\end{enumerate}
Step (2) and (3) are executed iteratively, until the desired goal is reached and the
distributed algorithm has completed its goal.

The initialization, local computation and aggregation functions can be customized to solve
different problems, while the framework takes care of providing the subgraph to
each worker, collecting the independent states from replicas and copying the aggregated state 
back to replicas.

To provide a couple of examples, Algorithms~\ref{alg:dist_init} and~\ref{alg:conn_init} show
how to compute the distances of vertices from a source vertex and to identify the connected components 
of a graph using our framework. Both are basic problems that can be used as building blocks
for other, more complex computations. For example, the problem of distance computation is
needed to compute properties like betweenness centrality~\cite{brandes2001}. It is also possible
to implement Luby's maximal independent set algorithm~\cite{luby1986} in \etsch, by spreading the
random values in the local phase and choosing if a vertex must be added to the set in the aggregation phase.

For the problem of distance computation (Algorithm~\ref{alg:dist_init}), each vertex is associated
with a state containing just the distance parameter $\Dist$. Initially, all
vertices are initialized to $+\infty$, apart from the \Source vertex which is
initialized to $0$. In the local computation phase, vertices are inserted in a
priority queue sorted by distance, from which are extracted to update the
distance of their neighbors (following the Dijkstra algorithm on the subgraph).
In the aggregation phase, replicated states of vertices are represented as 
a vector of distances, from which the minimum distance is taken.


\begin{algorithm}[t] 
\caption{Distance computation}
\label{alg:dist_init}
\FUNCTION{$\Init(\Node[\,]\ V_i)$}
{
	\ForEach{$v \in V$}
	{
	  \eIf{$v = \Source$}
	  {
	   $v.\Dist=0$\;
	   }{
	  $v.\Dist=\infty$\;
	  }
	}
}
\BlankLine
\FUNCTION{$\Local(\Node[\,]\ V_i)$}
{
	$PQ = \NEW\ \PriorityQueue \langle \Node \rangle()$\;
	\ForEach{$v \in V$}
	{ 
	  $PQ.\Add(v)$\;
	}
	\While{\NOT $PQ.\Empty()$}
	{
	  $u = PQ.\Pop()$\;
	  \ForEach{$v \in u.\Neighbors$}
	  {
	  	\If{$v.\Dist>u.\Dist+1$}
	  	{
	  		$v.\Dist=u.\Dist+1$\;
	  		$PQ.\Update(v)$\;
	  	}
	  }
	}
}
\BlankLine
\FUNCTION{$\Distance\ \Aggregation(\Distance[\,]\ D)$}
{
   \Return $\MIN(D)$\;
}
\end{algorithm}

\begin{algorithm}[t] 
\caption{Connected components computation}
\label{alg:conn_init}
\FUNCTION{$\Init(\Node[\,]\ V_i)$}
{
	\ForEach{$v \in G$}
	{
	   $v.\Id=\Random()$\;
	}
}
\BlankLine
\FUNCTION{$\Local(\Node[\,]\ V_i)$}
{
	$PQ = \NEW\ \PriorityQueue \langle \Node \rangle()$\;
	\ForEach{$v \in V$}
	{ 
	  $PQ.\Add(v)$\;
	}
	\While{\NOT $PQ.\empty()$}
	{
	  $q = PQ.\Pop()$\;
	  \ForEach{$v \in q.\Neighbors$}
	  {
	  	\If{$v.\Id>q.\Id$}
	  	{
	  		$v.\Id=q.\Id$\;
	  		$PQ.\Update(v)$\;
	  	}
	  }
	}
}
\BlankLine
\FUNCTION{$\Aggregation(\ID[\,]\ D)$}
{
   \Return $\MIN(D)$\;
}
\end{algorithm}

Computing the connected components works in a similar way (Algorithm~\ref{alg:conn_init}).
Each vertex is associated with a connected component identifier \Id, which is generated
randomly for each vertex. The local computation phase epidemically spread the smallest
component identifier by passing it through the local edges, until all vertices
have been reached. In the aggregation phase, the smallest
identifier is selected from all the replicas and returned as their connected component identifier.
Eventually, each connected component will be identified by a single value, which
is the smallest identifier randomly generated in each connected component.

\section{Distributed Funding-based Edge Partitioning}
\label{sec:algorithm}

The properties that a ``good'' partitioning must possess are the following:
\begin{itemize}
\item \textbf{Balance}: partition sizes should be as close as possible to the
average size $|E|/K$, where $K$ is the number of partitions. In this way, 
the amount of work needed in each partition is as
equal as possible.
\item \textbf{Communication efficiency}: given that the amount of communication
that crosses the border of a partition depends on the number of its frontier vertices,
the total sum $\sum_{i=1}^K |F_i|$ must be reduced as much
as possible.
\item \textbf{Connectedness}: the subgraphs induced by the partitions should be
connected. This is not a strict requirement (later in this section we
illustrate a variant of our algorithm that does not guarantee connected
partitions), but it allows us to see each subgraph as a completely independent
entity.
\item \textbf{Path compression}: a path between two vertices in $G$ is composed by
a sequence of edges. If some information must be passed across this path, it
will need to cross partitions every time two consecutive edges belong to
different partitions. The smallest the number of partitions to be traversed,
the better.
\end{itemize}

Balance is the main goal; note, however, that it would be simple to just split
the edges in $K$ sets of size $\approx |E|/K$,
but this could have severe implications on communication efficiency,
connectedness and path compression. The approach proposed here is thus
heuristic in nature and provides an approximate solution to the above
requirements.

Since the purpose is to compute the edge partitioning as a preprocessing step
to help the analysis of very large graphs, we need the edge partitioning
algorithm to be distributed as well. As with most distributed algorithms, we
are mostly interested in minimizing the amount of communication steps needed to
complete the partitioning.

Ideally, a simple solution could work as follows: to compute $K$ partitions,
$K$ edges are chosen at random and each partition grows around those edges.
Then, all partitions take control of the edges that are \emph{neighbors} (i.e.,
they share one vertex) of those already in control and are not taken by other
partitions. All partitions will incrementally get larger and larger until all
edges have been taken. Unfortunately, this simple approach does not work well
in practice, since the starting position may greatly influence the size of the
partitions. A partition that starts from the center of the graph will have more
space to expand than a partition that starts from the border and/or very close to
another partition.

To overcome this limitation, we introduce \dfep (Distributed Funding-based
Edge Partitioning), an algorithm based on concept of ``buying'' the edges
through an amount of \emph{funding} that is assigned to each partition.
Initially, each partition is assigned the same amount of funding and
an initial, randomly-selected vertex. The algorithm is then organized in a
sequence of \emph{rounds}. During each round, each partition makes an offer to
acquire the edges that are neighbors to those already taken. An edge is
then sold to the partition that makes the larger offer, and that partition has
to pay one \emph{unit} of funding. At the end of each round, a coordinator
monitors the sizes of each partition and sends additional units of funding.
Partitions that are smaller than average get more units of funding, to help them
overcome the slow start, while larger partitions receive only a small amount
of units. By tuning the amount of units given at the initialization
step and the amount of units sent during the execution it is possible to
obtain balanced partitions.

\begin{table}[h]
\caption{Notation \label{tab:notation}}
\centering
\begin{tabular}{|cc|}
\hline
 $d(v)$ & degree of vertex $v$ \\
 $E(v)$ & edges incident on vertex $v$ \\
 $V(e)$ & vertices incident on edge $e$ \\
 $M_i[v]$ & amount of units of partition $i$ in vertex $v$ \\
 $M_i[e]$ & amount of units of partition $i$ in edge $e$ \\
$E_i$ & edges bought by partition $i$ until now \\
 $\owner[e]$ & the partition that owns edge $e$ \\
\hline
\end{tabular}
\end{table}

Table~\ref{tab:notation} contains the notation used in the pseudocode of the
algorithm. For each vertex and edge we keep track of the amount of units that
each partition has committed to that vertex or edge. Algorithm~\ref{alg:init}
presents the code executed at the initialization step: each partition chooses a
vertex at random and assigns all the initial units to it. The edges are
initialized as unassigned. Each round of
the algorithm is then divided in three steps. In the first step (Algorithm
\ref{alg:step1}), each vertex propagates the units of funding to the outgoing edges.
For each partition, the vertex can move its funding only on edges that are free 
or owned by that partition, dividing the available units of funding equally 
among all these eligible edges. During the second step (Algorithm~\ref{alg:step2}), 
each free edge is bought by the partition which has the most units committed in that edge  and
the units of funding of the losing partitions are sent back in equal parts to the
vertices that contributed to that funding.
The winning partition loses an unit of funding to pay for the edge
and the remaining funding is divided in two equal parts and sent to the vertices composing the edge.
In the third step (Algorithm~\ref{alg:step3}), each
partition receives an amount of funding inversely proportional to the number of edges it
has already bought. This funding is distributed between all the vertices in
which the partition has already committed a positive amount of funding.

\begin{figure}[b]
\begin{center}
\includegraphics[width=.15\textwidth]{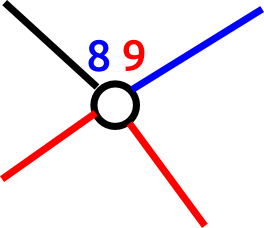}
\qquad
\includegraphics[width=.15\textwidth]{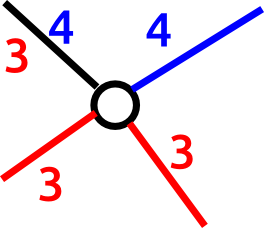}
\end{center}
\caption{Example execution of \dfep, step 1.}
\label{fig:ex-node}
~\\
%
\begin{center}
\includegraphics[width=.15\textwidth]{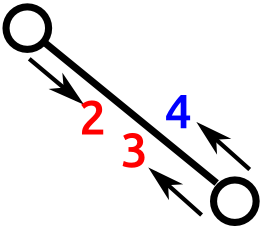}
\qquad
\includegraphics[width=.15\textwidth]{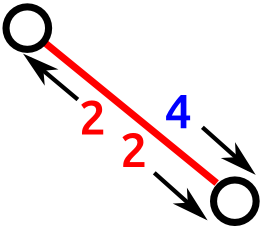}
\end{center}
\caption{Example execution of \dfep, step 2.}
\label{fig:ex-edge}
\end{figure}

A couple of examples are illustrated in
Figures~\ref{fig:ex-node}-\ref{fig:ex-edge}. The red and blue color represents
partitions, while black edges are still free. 
Figure~\ref{fig:ex-node} illustrates step 1 of the algorithm. The
vertex has $8$ units on the blue partition, $9$ units on the red one. Two edges
are owned by the red partition, one by the blue, and the black one is still
unassigned. When step 1 is concluded, the $9$ red units have been committed to
the two red edges and the black one, while the $8$ blue units have been committed
to the blue edge and the black one. The blue partition will be be allowed to
buy the black edge. Figure~\ref{fig:ex-edge} illustrates step 2 executed on a
single edge. The edge receives $5$ red units and $4$ blue units, and thus is
assigned to the red partition. All the blue units are returned to the sender
while the remaining $5-1$ red units are divided equally between the two vertices.

\begin{algorithm}[t]
\caption{\dfep Init: executed by coordinator}
\label{alg:init}
\ForEach{edge $e \in E$}
{
	$\owner[e]=\bot$\;
}
\For{$i = 1$ \TO\ $K$}
{ 
  $v \gets random(V)$\; 
  $M_i[v] = |E| / K $\;
}
\end{algorithm}

\begin{algorithm}[t]
\caption{\dfep Step 1: executed at each vertex $v$}
\label{alg:step1}
\For{$i = 1$ \TO\ $K$}
{
  \If{$M_i[v]>0$}
  {
    $\eligible=\emptyset$\;
    \ForEach{$e \in E(v)$}
    {
      \If{$\owner[e]=\bot$ \OR $\owner[e]=i$}
      {
        $\eligible=\eligible \cup \{e\}$\;
      }
    }
    \ForEach{$e \in \eligible$}
    {
      $M_i[e]=M_i[e] + (M_i[v] / |\eligible|)$\;
    }
    $M_i[v]=0$\;
  }
}
\end{algorithm}

\dfep creates partitions that are connected subgraphs of the original graph,
since the funding cannot traverse an edge that has not been bought by that
partition. It can be implemented in a distributed framework: both step 1 and
step 2 are completely decentralized; step 3, while centralized, needs an
amount of computation that is only linear in the number of partitions.

In our implementation, the amount of initial funding is equal 
to what would be needed to buy an amount of edges equal to the 
optimal sized partition. A smaller quantity would not decrease the precision of 
the algorithm, but it would slow it down during the first rounds. 
The cap on the units of funding to be given to a small partition during each round (10 in our 
implementation) avoids the overfunding of a small partition during the first rounds.

\begin{algorithm}[t]
\caption{\dfep Step 2: executed at each edge $e$}
 \label{alg:step2}
$best= argmax_p(M_p(e))$\;
\If{$\owner[e]=\bot$ \AND $M_{best}(e)\geq1$}
{
  $\owner[e]=best$\;
  $M_{best}[e] = M_{best}[e]-1$\;
}
\For{$i = 1$ \TO\ $K$}
        {
          \eIf{$\owner[e]=i$}
              {
                \ForEach{$v \in N(e)$}
                        {
                          $M_i[v] = M_i[v] + M_i[e]/2$
                        }
              }{
                $S=$ vertices that funded partition $i$ in $e$\;
                \ForEach{$v \in S$}
                        {
		          $M_i[v] = M_i[v] + M_i[e]/|S|$\;
                        }
              }
              $M_i[e]=0$\;
        }
\end{algorithm}

\begin{algorithm}[t]
\caption{\dfep Step 3: executed by the coordinator}
\label{alg:step3}
$AVG= \sum_{i \in [1 \ldots K]}(|E_i|) / K$\;
\For{$i = 1$ \TO\ $K$}
{
  $\funding= min(10, AVG / E_i) $\;
  \ForEach{$v \in V$}
          {
            \If{$M_i(v)> 0$}
               {
                 $M_i(v)= M_i(v) + \funding$\;
               }
          }
}
\end{algorithm}

\subsection{\dfep Variant}
If the diameter is very large, there is the possibility that a poor starting vertex is chosen at the beginning 
of the round. A partition may be cut off from the rest of the graph, thus creating unbalanced 
partitions. A possible solution for this problem involves adding an additional dynamic, at the cost of 
losing the connectedness property. 

A partition is called \emph{poor} at round $i$ if its size is less  than $\frac{\mu}{p}$, with $\mu$ being 
the average size of partitions at round $i$ and $p$ being an additional parameter; otherwise, it is called
\emph{rich}. A poor partition can commit units on already 
bought edges that are owned by rich partitions and try to buy them. This addition to the algorithm allows 
small partition to catch up to the bigger ones even if they have no free neighboring edges and results in
more balanced partitions.

\section{Results} \label{sec:results}

This section starts by introducing the different metrics that have been
measured during the experiments and the datasets that have been used. Then, the
evaluation is split in two parts: first, using a simulation engine, we evaluate
in detail the behavior of \dfep; then, using the Amazon EC2 cluster, we
evaluate whether \etsch actually improves the computing time of the shortest
path algorithm introduced in Section~\ref{sec:framework}.

\subsection{Metrics}

In the Amazon EC2 cluster, the most important metric is the actual running time
of our algorithm and how does it scale with the number of machines. The simulation
engine, on the other hand, allows us to obtain a better understanding of the
behavior of \dfep; the metrics considered in such case are the following:

\metric{Number of rounds} the number of rounds executed by \dfep to complete
the partitioning. This is a good measure of the amount of synchronization
needed and can be a good indicator of the eventual running time in a real world
scenario.

\metric{Balance} Each partition should be as close as possible to the same 
size. To obtain a measure the balance between the partitions we first normalize 
the sizes, so that a partition of size 1 represent a partition with exactly $|E| / 
K$ edges. We then measure both the size of the largest partition and the standard 
deviation of the sizes, computed as in the following formula. $E$ is the number of 
vertices, $K$ is the number of partitions and $|E_i|$ is the size of the $i$-th 
partition):
$$\mathit{NSTDEV} = \sqrt{ \dfrac{\sum_{i=1}^K{(\dfrac{|E_i|}{E/K}-1)^2}}{K}}$$

\metric{Communication cost} As illustrated in Section~\ref{sec:framework}, at
the end of each round all vertices that appear in multiple partitions must
collapse their state to a common value. The amount of messages needed by \etsch
when executed on a specific edge partitioning is computed using the following
formula ($F_i$ is the set of frontier vertices of partition $i$).
$$\mathit{MESSAGES} = \sum_{i=1}^K {F_i}$$

\metric{Path compression} A good edge-partitioning will also reduce the number
of rounds needed by \etsch to finish its computation. How much will it improve
\etsch performances depends on the specific problem, therefore we chose the
shortest path algorithm presented in Section~\ref{sec:framework} as a
representative. We thus call the \emph{gain} of an edge-partitioning of a graph
the fraction of total iterations avoided by the shortest path algorithm implemented
in \etsch.

\subsection{Datasets}

Since the simulation engine is not able to cope with larger datasets, we used
different datasets for the experiments in the simulation engine and  the
real world experiments. For both types of datasets we list the size of the
graphs, the diameter $D$, the clustering coefficient $CC$ and the
clustering coefficient $RCC$ of a random graph with the same size.

Table~\ref{tab:datasets_sim} contains the characteristics of the four different
datasets used in the simulation engine. \dataset{astroph} is a collaboration
network in the astrophysics field, while \dataset{email-enron} is an email
communication network from Enron. Both datasets are small-world, as shown by
the small diameter. The \dataset{usroads} dataset is a road networking the US, and
thus is a good example of a large diameter network. Finally, \dataset{wordnet} is
a synonym network, with small diameter and very high clustering coefficient.

The three larger graphs that are used to run the Hadoop implementation of both
\dfep and \etsch are presented in Table~\ref{tab:datasets_exp}. \dataset{dblp} is
the co-authorship network from the DBLP archive, \dataset{youtube} is the
friendship graph between the users of the service while \dataset{amazon} is a
co-purchasing network of the products sold by the website.

All the networks have been taken from the SNAP graph library~\cite{Leskovec11}
and cleaned for our use, making directed edges undirected and removing
disconnected components.

\begin{table}
\centering
\caption{Datasets used in the simulation engine \label{tab:datasets_sim}}
\begin{tabular}{|c|ccccc|}
\hline
Name & $|V|$ & $|E|$ & $D$ & $CC$ & $RCC$\\
\hline
\dataset{astroph} & 17903 & 196972 & 14 & $1.34\times10^{-1}$ & $1.23\times10^{-3}$\\
\dataset{email-enron} & 33696 & 180811 & 13 & $3.01\times10^{-2}$ & $3.19\times10^{-4}$ \\
\dataset{usroads} & 126146 & 161950 & 617 & $1.45\times10^{-2}$ &$2.03\times10^{-5}$ \\ 
\dataset{wordnet} & 75606 & 231622 & 14 & $7.12\times10^{-2}$ & $8.10\times10^{-5}$\\
\hline
\end{tabular}
\vspace{0.5cm}
%
\caption{Datasets used on the Amazon EC2 cloud \label{tab:datasets_exp}}
\centering
\begin{tabular}{|c|ccccc|}
\hline
Name & $|V|$ & $|E|$ & $D$ & $CC$ & $RCC$\\
\hline
\dataset{dblp} & 317080 & 1049866 & 21 & $1.28\times10^{-1}$& 2.09$\times10^{-5}$\\
\dataset{youtube} & 1134890 & 2987624 & 20 & $2.08 \times10^{-3}$& $4.64\times10^{-6}$ \\
\dataset{amazon} & 400727 & 2349869 & 18 & $5.99\times10^{-2}$& $2.93\times10^{-5}$\\ 
\hline
\end{tabular}
\end{table}

\subsection{Simulations}

\begin{figure*}[!ht]
\setlength{\belowcaptionskip}{-5pt}
\captionsetup[subfigure]{skip=-12pt}

\begin{subfigure}[b]{0.45\textwidth}
\includegraphics[width=\textwidth]{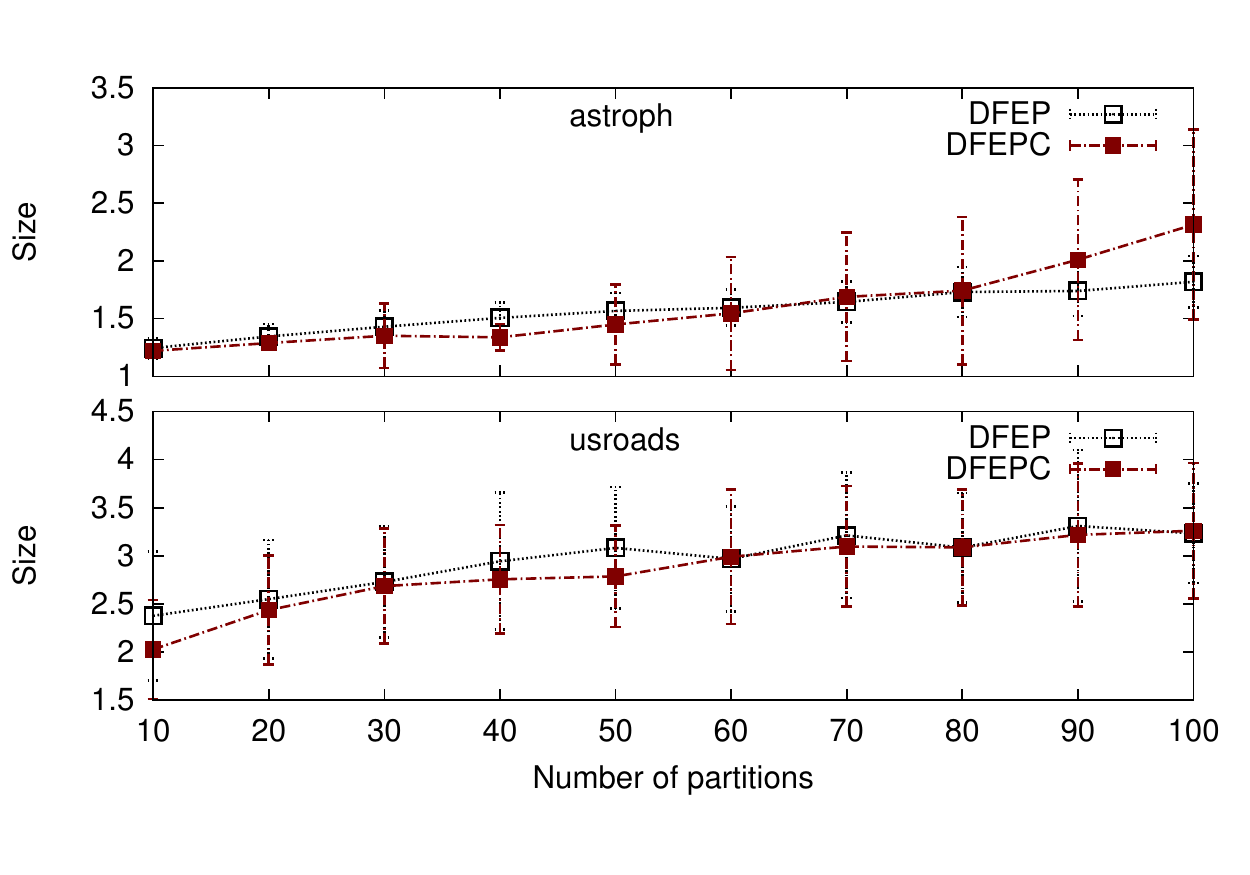}
\caption{Size of the largest partition}
\label{fig:growingK-maxsize}
\end{subfigure}
\hfill
\begin{subfigure}[b]{0.45\textwidth}
\includegraphics[width=\textwidth]{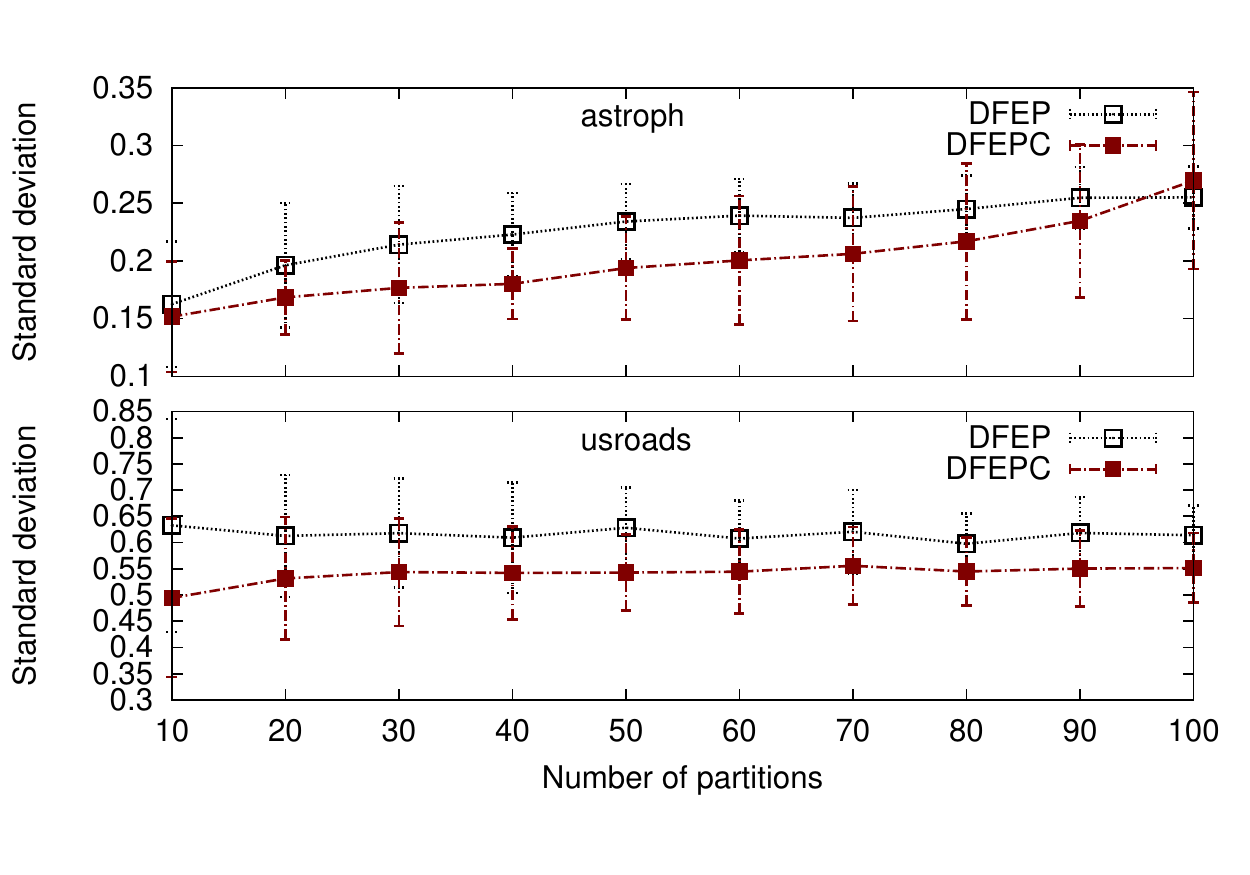}
\caption{Standard deviation of partition sizes}
\label{fig:growingK-stdev}
\end{subfigure}

\vspace{-5pt}
\begin{subfigure}[b]{0.45\textwidth}
\includegraphics[width=\textwidth]{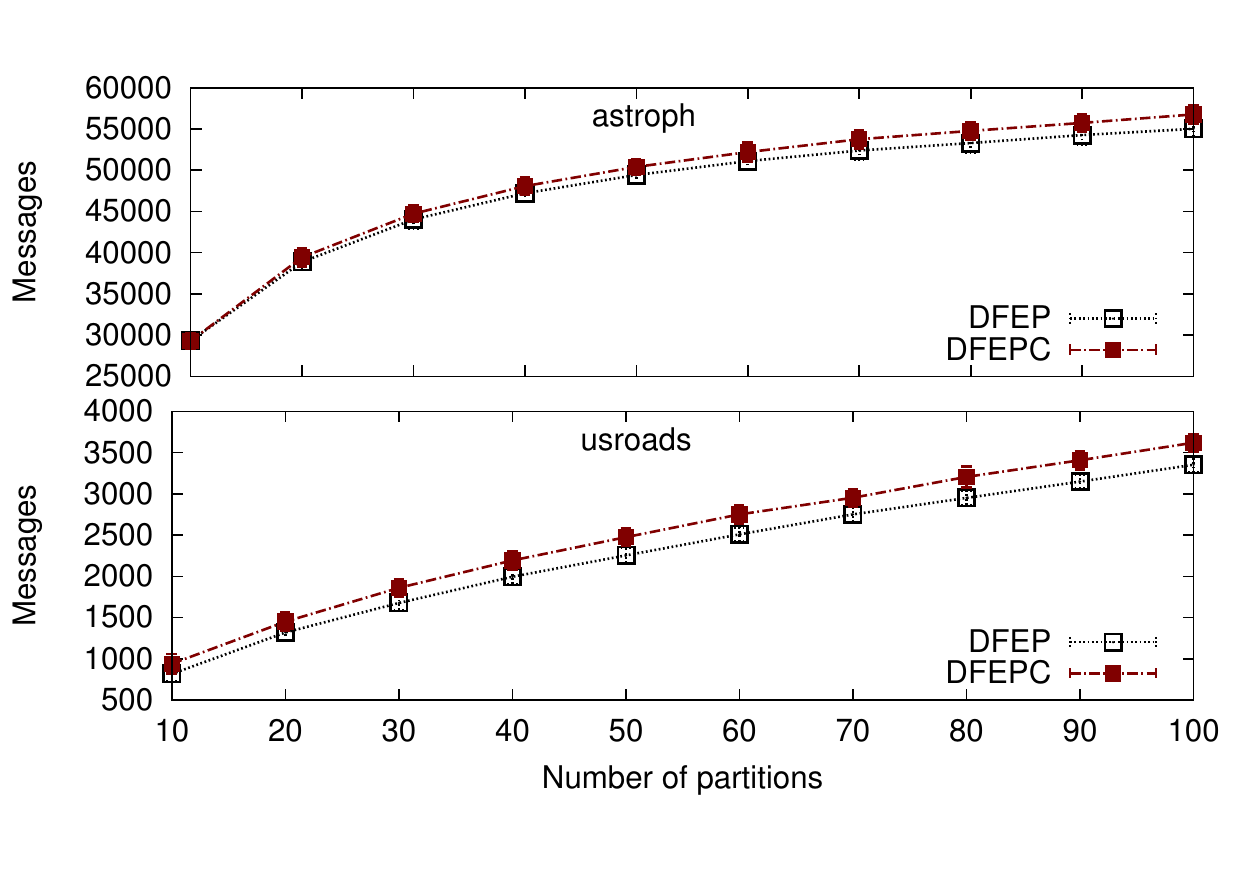}
\caption{Communication cost}
\label{fig:growingK-mess}
\end{subfigure}
\hfill
\begin{subfigure}[b]{0.45\textwidth}
\includegraphics[width=\textwidth]{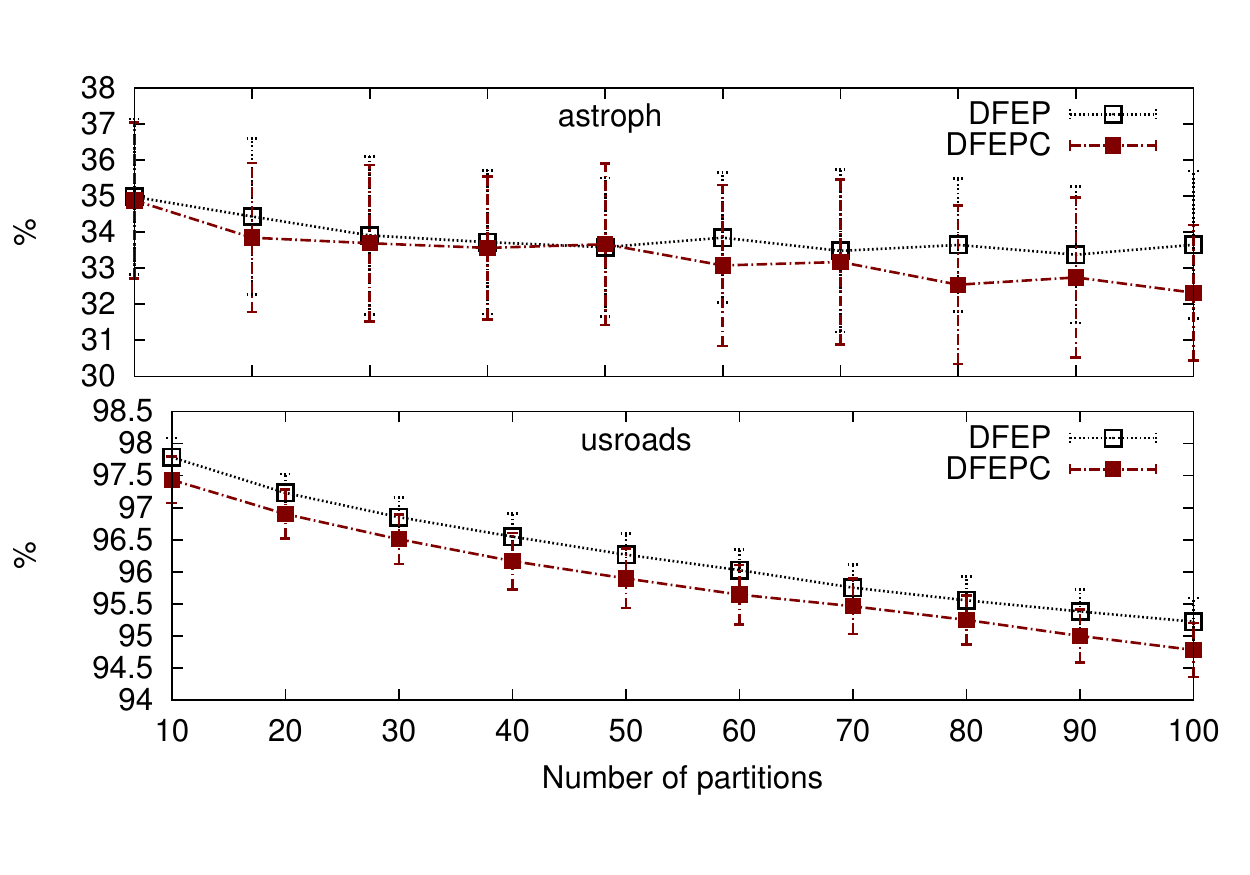}
\caption{Gain on iterations}
\label{fig:growingK-adv}
\end{subfigure}

\vspace{-5pt}
\begin{subfigure}[b]{0.45\textwidth}
\includegraphics[width=\textwidth]{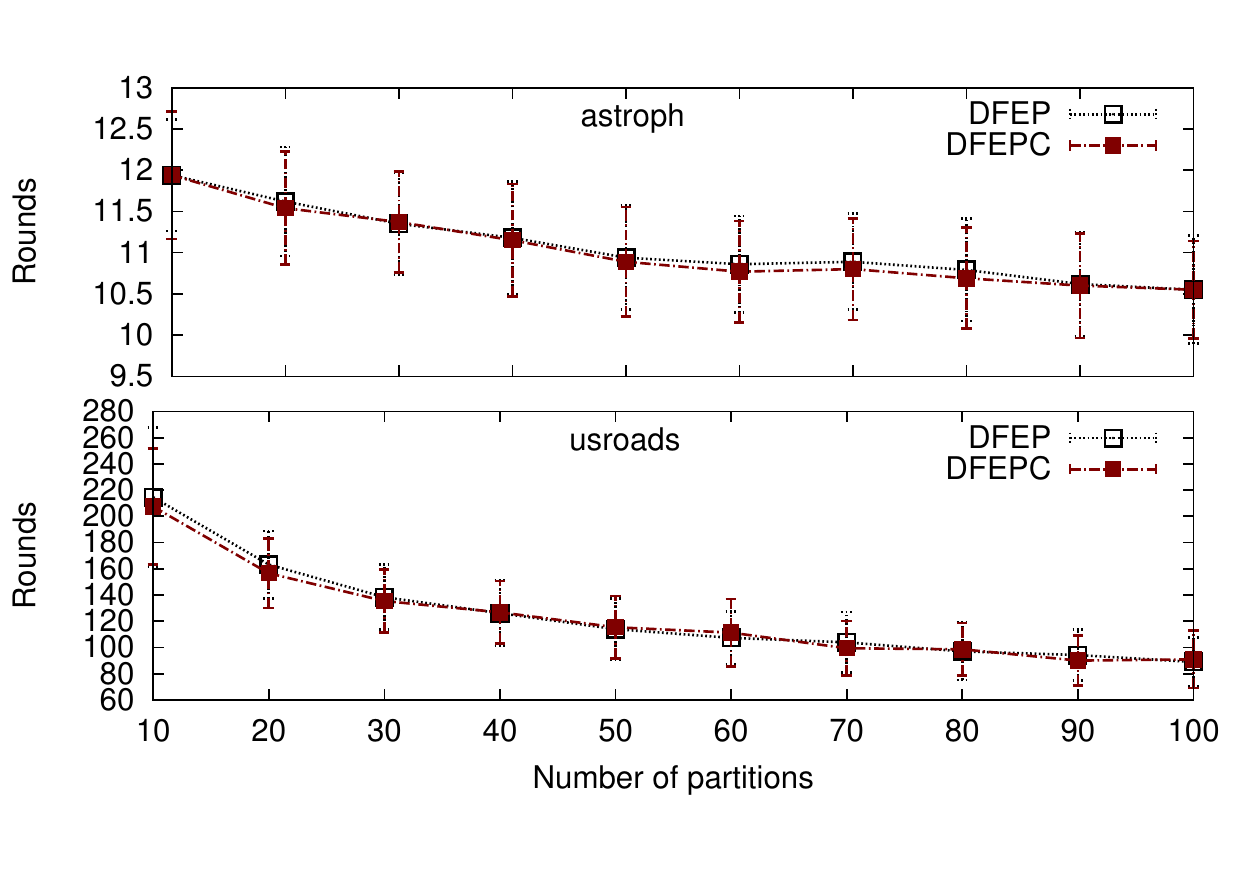}
\caption{Rounds needed by \dfep to converge}
\label{fig:growingK-iter}
\end{subfigure}
\hfill
\begin{subfigure}[b]{0.45\textwidth}
\includegraphics[width=\textwidth]{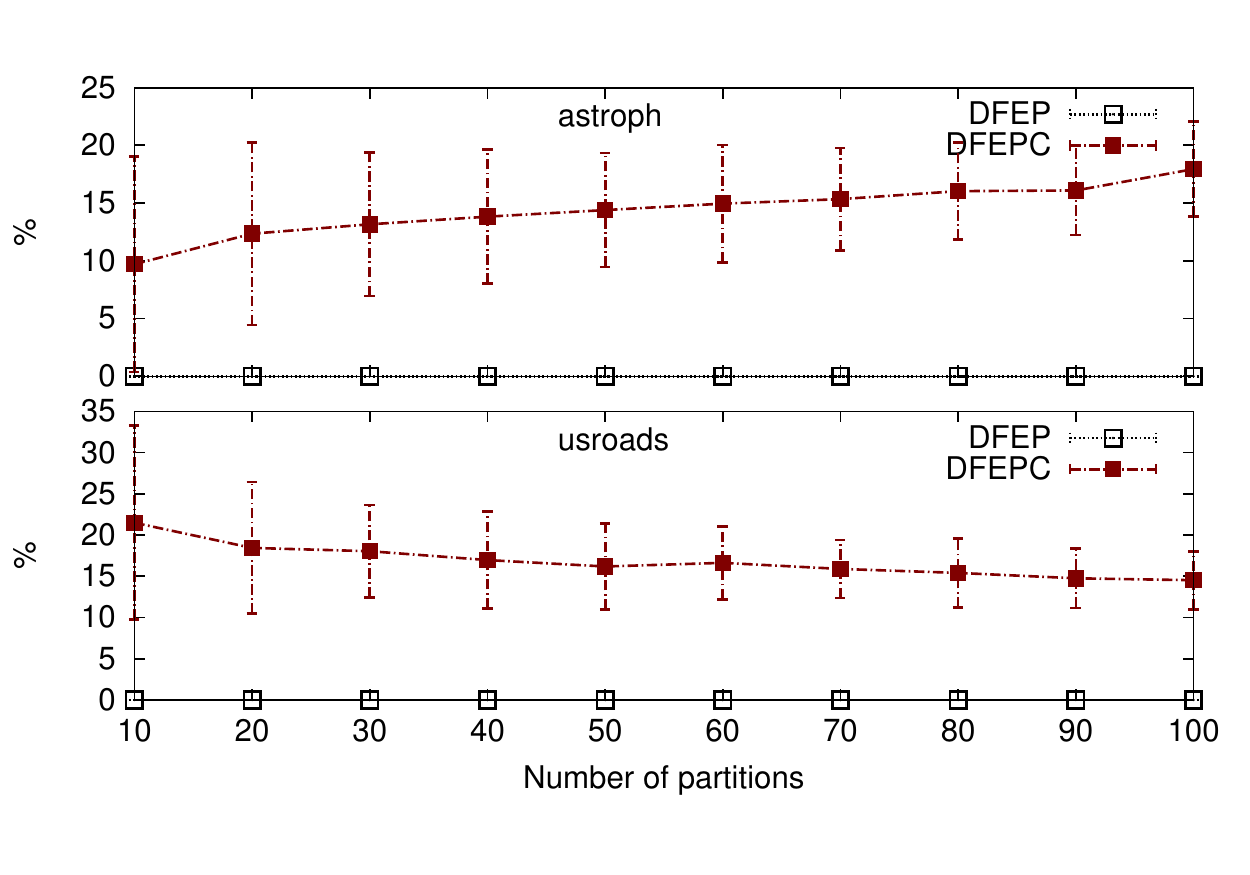}
\caption{Percentage of disconnected partitions}
\label{fig:growingK-conn}
\end{subfigure}

\caption{Behavior of \dfep and \dfepc with different values of $K$ ($100$ samples)}
\label{fig:growingK}
\end{figure*}

Figure~\ref{fig:growingK} shows the performance of the two versions of
\dfep against the parameter $K$, in the \dataset{astroph} and
\dataset{usroads} datasets. As expected, the larger the number of partitions, the larger
is the variance between the sizes of those partitions and the amount of
messages that will have to be sent across the network. The rounds needed to
converge to a solution go down with the number of partitions, since it will
take less time for the partitions to cover the entire graph. Finally, the gain
obtained by using \etsch is larger when there are only few partitions, since
the paths are more compressed. This property will emerge also from the
experimental results on the EC2 cloud.

The diameter of a graph is a strong indicator of how our proposed approach will 
behave. To test \dfep on graphs with similar characteristics but different diameter 
we followed a specific protocol: starting from the \dataset{usroads} dataset (a graph with 
a very large diameter) we remapped random edges, thus decreasing the diameter. The
remapping has been performed in such a way to keep the number of triangles as close
as possible to the original graph. 

\begin{figure*}
\setlength{\belowcaptionskip}{-5pt}

\begin{center}
\begin{subfigure}[b]{0.3\textwidth}
\includegraphics[width=\textwidth]{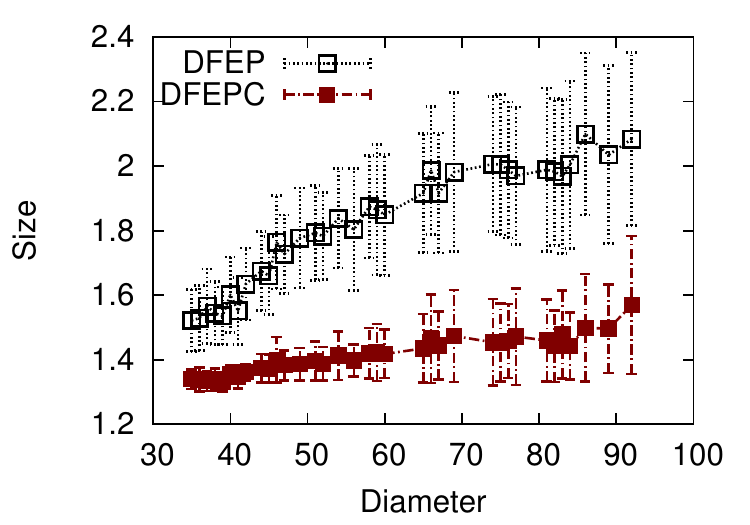}
\caption{Relative size of largest partition}
\label{}
\end{subfigure}
\begin{subfigure}[b]{0.3\textwidth}
\includegraphics[width=\textwidth]{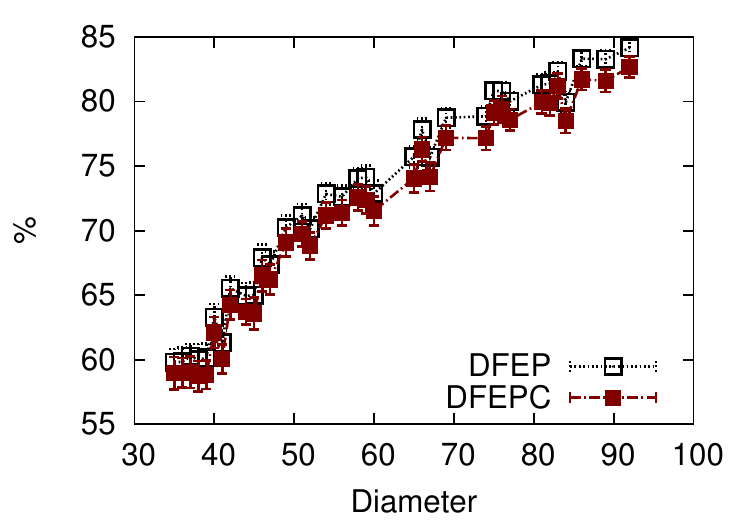}
\caption{Gain on iterations by \etsch}
\label{}
\end{subfigure}
\begin{subfigure}[b]{0.3\textwidth}
\includegraphics[width=\textwidth]{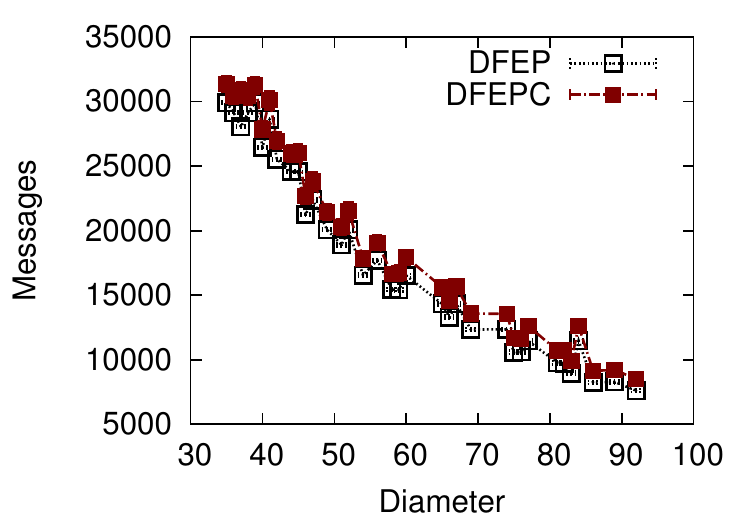}
\caption{Communication cost}
\label{}
\end{subfigure}
\end{center}

\vspace{-10pt}
\begin{center}
\begin{subfigure}[b]{0.3\textwidth}
\includegraphics[width=\textwidth]{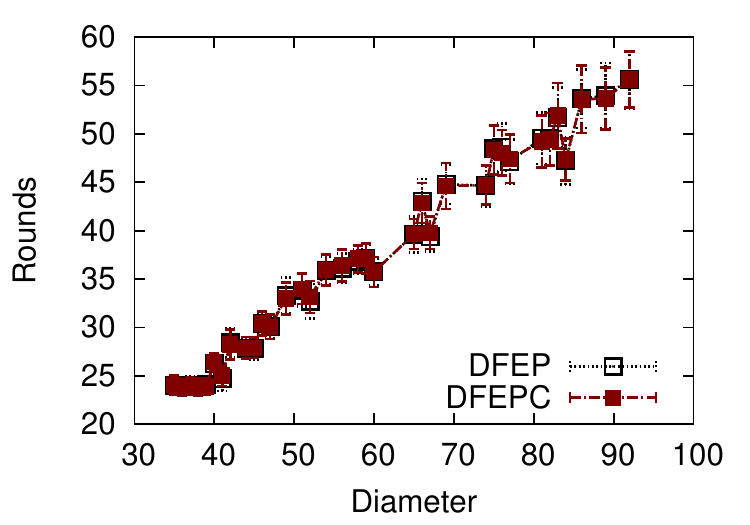}
\caption{Rounds needed by \dfep to converge}
\label{}
\end{subfigure}
\begin{subfigure}[b]{0.3\textwidth}
\includegraphics[width=\textwidth]{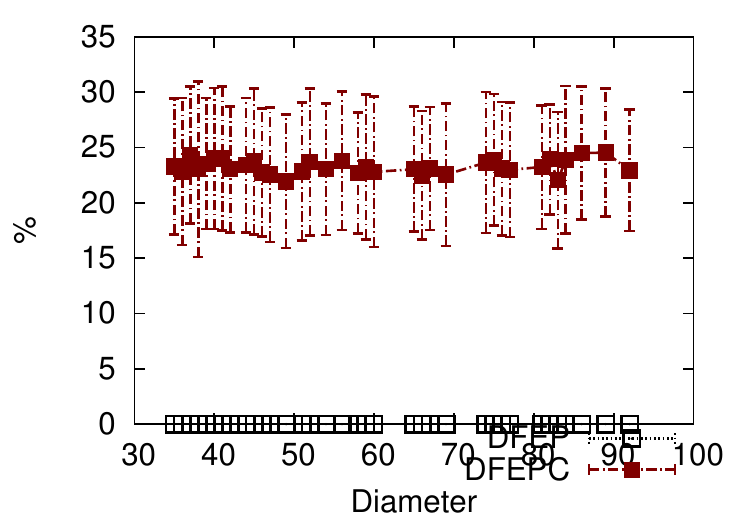}
\caption{Percentage of disconnected partitions}
\label{}
\end{subfigure}
\begin{subfigure}[b]{0.3\textwidth}
\includegraphics[width=\textwidth]{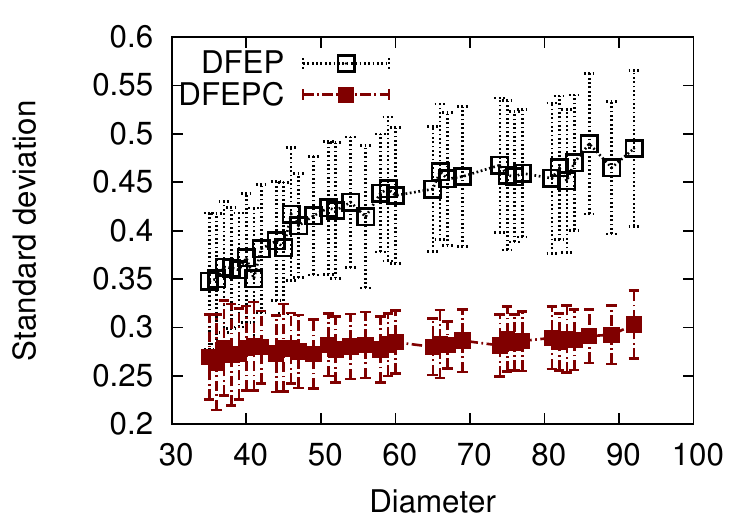}
\caption{Standard deviation of partition sizes}
\label{}
\end{subfigure}
\end{center}

\caption{Behavior of \dfep with graphs of same size but different diameter ($K=20$, $100$ samples)}
\label{fig:diameter}
\end{figure*}

Figure~\ref{fig:diameter} shows that changing the diameter leads to completely
different behaviors. The size of the largest partitions and the standard
deviation of partitions size rise steeply with the growth of the diameter,
since in a graph with higher diameter the starting vertices chosen by our
algorithm affect more deeply the quality of the partitioning. As expected, the
number of rounds needed also rise linearly with the diameter, as does the gain
of \etsch. While the partitions may be less balanced, they will be more
interconnected and thus the amount of messages sent across the network will
decrease steeply.

Finally, we compare the 
two version of \dfep against the JaBeJa\cite{Rahimian2013} algorithm. Since JaBeJa is a vertex-
partitioning algorithm, its output must be converted into an edge-partitioning. 
Two approaches have been considered: running the algorithm directly on the line 
graph of the input graph, creating a vertex for each edge in the original graph, or 
assigning each edge to a partition by following the vertex-partitioning and 
assigning each cut edge randomly to one of the two neighboring partitions. Since 
the line graph can be orders of magnitude bigger than the original graph we followed the second approach.

\begin{figure*}
\setlength{\belowcaptionskip}{-5pt}

\begin{center}
\begin{subfigure}[b]{0.3\textwidth}
\includegraphics[width=\textwidth]{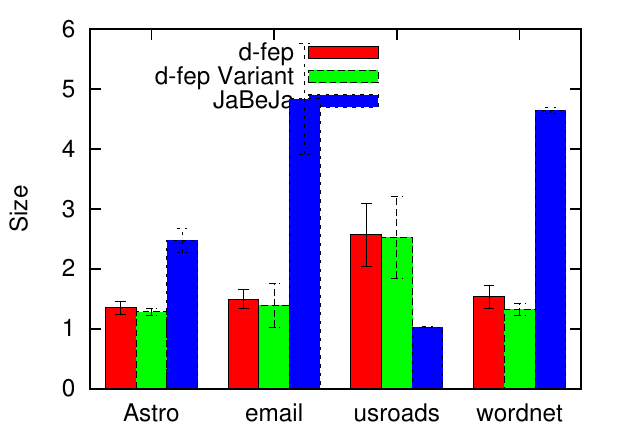}
\caption{Relative size of largest partition}
\label{}
\end{subfigure}
\begin{subfigure}[b]{0.3\textwidth}
\includegraphics[width=\textwidth]{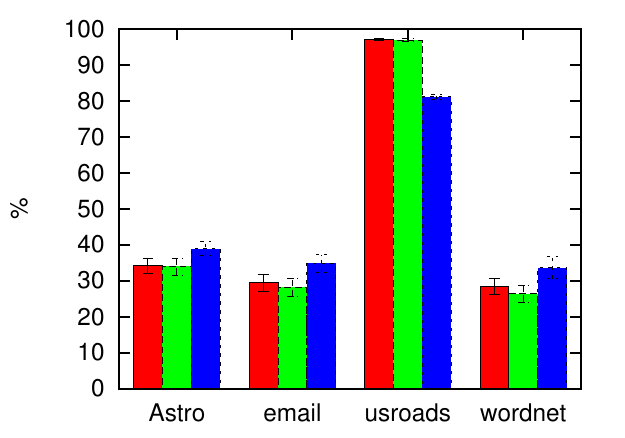}
\caption{Gain on iterations by \etsch}
\label{}
\end{subfigure}
\begin{subfigure}[b]{0.3\textwidth}
\includegraphics[width=\textwidth]{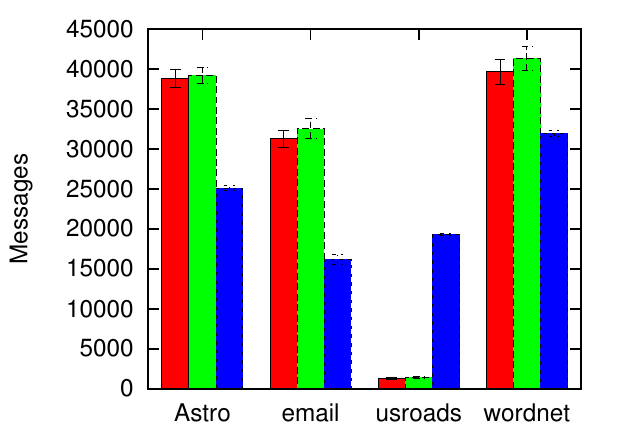}
\caption{Communication cost}
\label{}
\end{subfigure}
\end{center}

\vspace{-8pt}
\begin{center}
\begin{subfigure}[b]{0.3\textwidth}
\includegraphics[width=\textwidth]{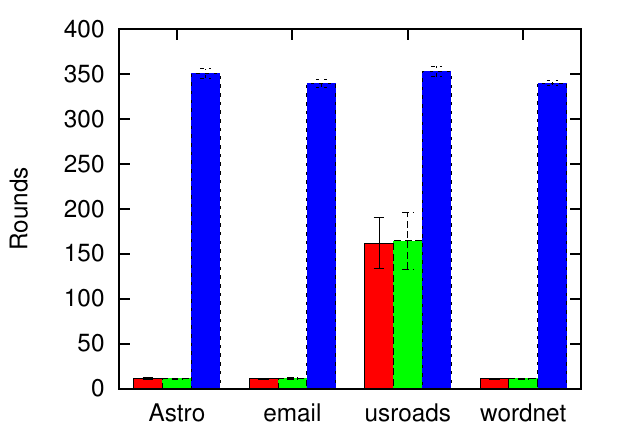}
\caption{Rounds needed to converge}
\label{}
\end{subfigure}
\begin{subfigure}[b]{0.3\textwidth}
\includegraphics[width=\textwidth]{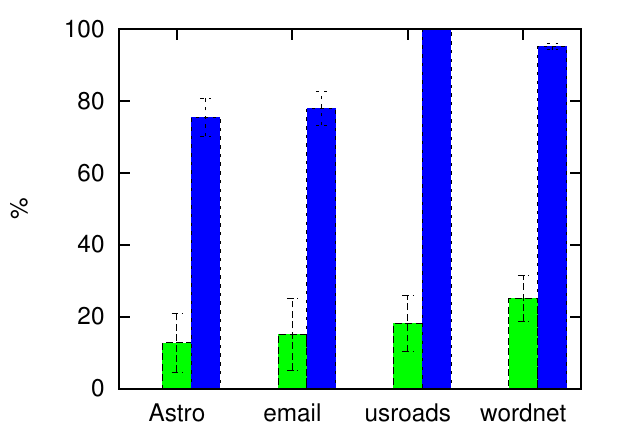}
\caption{Percentage of disconnected partitions}
\label{}
\end{subfigure}
\begin{subfigure}[b]{0.3\textwidth}
\includegraphics[width=\textwidth]{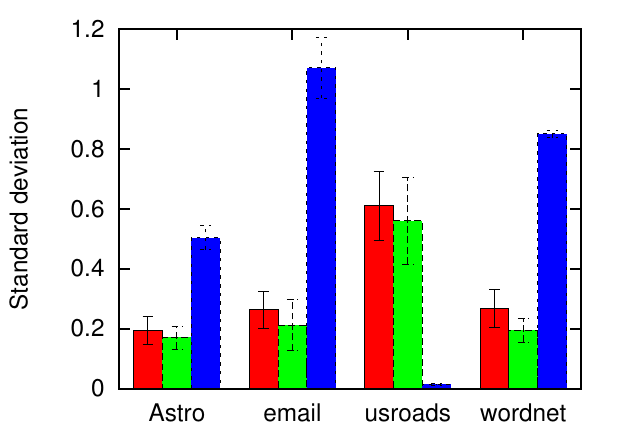}
\caption{Standard deviation of partition sizes}
\label{}
\end{subfigure}
\end{center}

\caption{Comparison between the two versions of \dfep and JaBeJa ($K=20$, $100$ samples)}
\label{fig:comparison}
\end{figure*}

Figure \ref{fig:comparison}
shows the experimental results over 100 samples, on the four different datasets.
A pattern can be discerned: the algorithms have wildly different behaviors in the 
small world dataset than in the road network. In the small world datasets our 
approaches results in more balanced partitions, while keeping the gain similar to 
JaBeJa. In the \dataset{usroads} dataset JaBeJa creates more balanced
partitions, but the gain is much lower and, more importantly, the amount of
messages that have to been sent is roughly ten times higher. This result shows
the importance of creating partitions that are as much connected as possible.

Since JaBeJa uses simulated annealing to improve the candidate solution, the
number of round needed is mostly independent from the structure of the graph.
As shown in Figure \ref{fig:diameter} the number of rounds \dfep needs depend mostly
from the diameter of the graph.

\subsection{Amazon EC2}

Both \dfep and \etsch have been also implemented in Apache Hadoop, in the
MapReduce model and tested over the Amazon EC2 cloud. All the experiments have
been repeated $20$ times on \emph{m1.medium} machines initiated by the Apache
Whirr tool, using the version 1.2.1 of Hadoop.

In the \dfep implementation the $K$ edges from which the partitions should start are 
chosen using a simple selection algorithm: each edge computes a random number in 
the Map phase and through first the usage of Combiners and finally of a single 
Reducer the $K$ edges that chose the smallest $K$ numbers are selected and assigned to 
a single partition each. 
It was not possible to implement \dfep using a single Map-Reduce round for 
each iteration while keeping exactly the same structure. Each instance of the Map 
function is executed on a single vertex, which will output messages to its 
neighbor and a copy of itself. Each instance of the Reduce function will receive 
a vertex and all the funding sent by the neighbors on common edges. The part of 
the algorithm that should be executed on each edge is instead executed by both its 
neighboring vertices, with special care to make sure that both executions will get 
the same results to avoid inconsistencies in the graph. This choice, which sounds 
counterintuitive, allows us to use a single Map-Reduce round for each iteration of the algorithm, 
thus decreasing the communication and sorting costs inherent in the MapReduce model.

\begin{figure}  
\includegraphics[width=.45\textwidth]{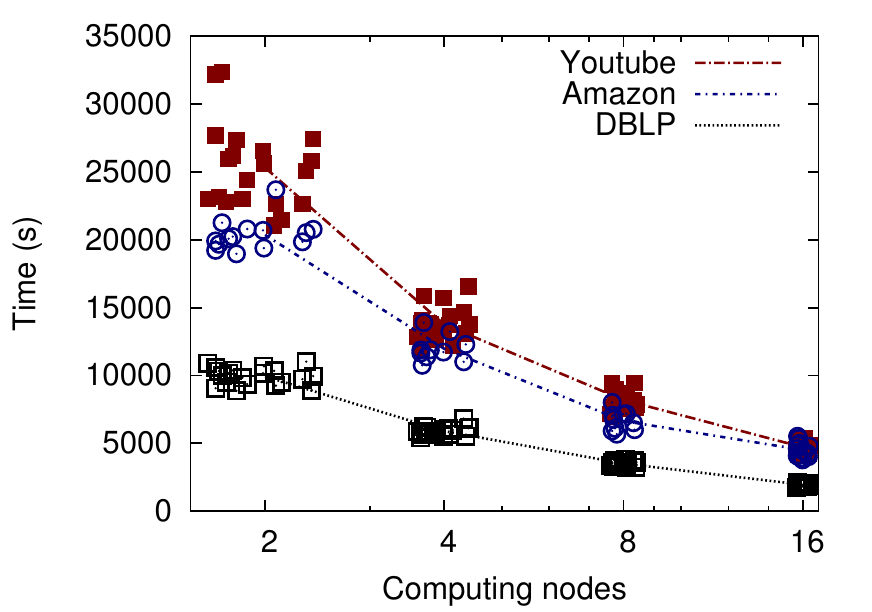}
\caption{Speedup of \dfep in the amazon cloud}\label{fig:dfep_amazon}
\end{figure}

Figure~\ref{fig:dfep_amazon} presents the scalability results for the \dfep
algorithm, when run with the three different datasets that are listed in
Table~\ref{tab:datasets_exp}, with $K=20$. The algorithm scales with the number
of computing nodes, with a speedup larger than $5$ with $16$ nodes instead of $2$.

\begin{figure} 
\includegraphics[width=.45\textwidth]{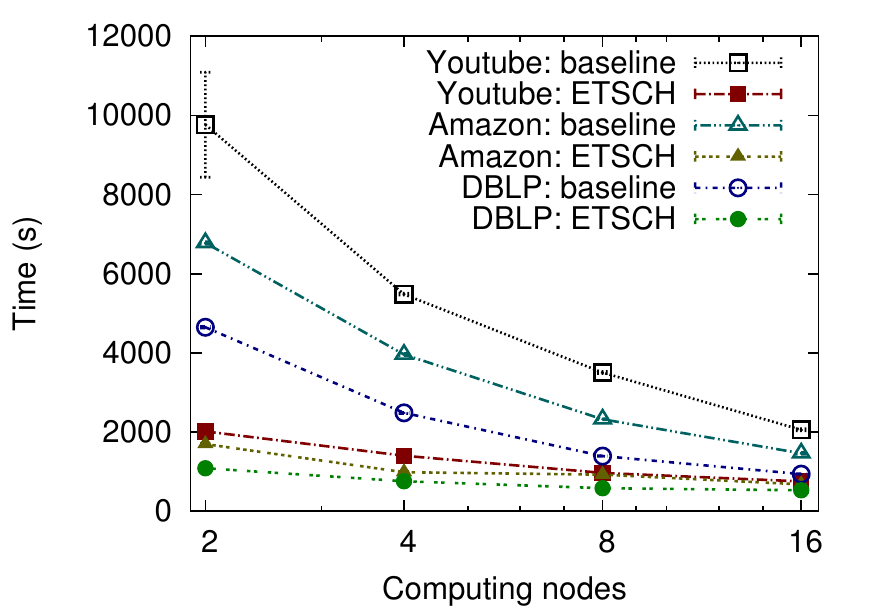}
\caption{Running time of single source shortest path algorithm using a standard
baseline algorithm and using our partitioned graph and the processing framework
introduced in Section \ref{sec:framework}}\label{fig:comp_amazon}
\end{figure}

To test the practical advantages of \etsch we prepared a Hadoop implementation
of the framework in which the user can define the three functions as defined in
Section~\ref{sec:framework}. We used the edge-partitioning obtained by running
\dfep, setting the number of desired partitions equal to the number of
available nodes, and run the \etsch implementation of the shortest path
algorithm. We compare this approach against running our baseline vertex-based
implementation of the shortest path algorithm on the unpartitioned graph.
Figure~\ref{fig:comp_amazon} shows that our approach is much more efficient
when the number of processing nodes is small, since the partitions are larger
and paths are more easily compressed. When the number of partitions grows, the
baseline approach gets closer to \etsch, but still seems less efficient.

While the baseline implementation could easily be optimized, the same can be
said about our implementation of the \etsch framework.
The experimental results thus show that \etsch and edge-partitioning is a 
promising approach.

\section{Related work}
\label{sec:stateart}
This section  is organized in two parts: first we introduce the most used 
frameworks for distributed graph analysis, briefly discussing their pros and 
cons. The second part presents and compare several known approaches for graph 
partitioning.

\subsection{Distributed frameworks for data analysis}
\label{sec:frameworks}

The MapReduce programming model~\cite{Dean2008} has been introduced by Google
to facilitate the development and execution of algorithms on very large quantities
of data. This model
inherits the map and reduce functions from functional programming to create a
simple and inherently parallelizable programming model. 
While the MapReduce programming model has been proposed by Google, the most common
open-source implementation is Apache's Hadoop~\cite{Bialecki2005}.

While one of the original examples of MapReduce application was PageRank~\cite{Dean2008}, 
the programming model is not very efficient for
graph analysis. The graph structure has to be passed across the map and
reduce functions and the entire graph must be read and rewritten at each iteration
whenever an iterative execution of the MapReduce paradigm is needed.

Pregel~\cite{Malewicz2010} was developed again by Google as an answer to these issues.
In similar vein to the Bulk Synchronous
Parallel model~\cite{Valiant1990}, each iteration is composed of two phases,
computation and communication, terminated by a single synchronization barrier. Each
vertex is represented as a process, with knowledge of its own neighbors.
In the first phase, the processes independently execute a number of computation steps
 and possibly issue messages to other processes. During the
second phase, all the messages are sent across the network and delivered to the
processes. The synchronization barrier makes sure that all vertices receive
all messages sent to them during the current iteration before the start of the
next one.  During the
computation phase each process updates its state by using the messages arrived
during the previous iteration, sends messages to its neighbors and may vote
to halt the computation. If all vertices vote to terminate, all
processes are stopped and the output is written to disk.

A different approach is offered by the GraphLab framework~\cite{Low10}, with the 
asynchronous
Gather-Apply-Scatter pattern. In the Gather phase each vertex receives the states of 
neighboring
vertices, changes to the local states are implemented in the Apply phase and eventual 
changes
are spread across outgoing edges in the Scatter phase.

\subsection{Graph Partitioning}

The literature on the graph partitioning problem is huge, but given that edge partitioning 
has not been studied in equal depth, we will focus on the different approaches developed to 
solve vertex graph partitioning. The edge partitioning problem can be
reduced to the vertex partitioning problem by using the line graph of the original graph, but the
massive increase in size makes this approach unfeasible.

In both versions, the partitioning problem is not only NP-Complete, but even
difficult to approximate~\cite{Andreev2004}. Most work in this field are thus
heuristics algorithms with no guaranteed approximation rate. Kernighan and Lin
developed the most well-known heuristic algorithm for binary graph partitioning
in 1970~\cite{Kernighan1970}. At initialization time, each vertex in the network is randomly assigned to one 
of the two partitions and the algorithm tries to optimize the vertex cut by exchanging vertices between the 
partitions. This approach has been later extended to run efficiently on multiprocessors by 
parallelizing the computation of the scoring function used to choose which vertices should be exchanged~\cite{Gilbert1987}.

METIS~\cite{Karypis1995} is a more recent and highly successful project that uses a multilevel 
partitioning approach to obtain very high quality partitions. The graph is coarsened into a smaller
graph, which is then partitioned and the solution is then refined to adapt to the original graph.
An effort to create a parallelizable version of the program has lead to P-METIS,
a version built for multicore machines. The quality of the partitions obtained with this approach 
does not seem to be of the same quality than the centralized version, as expected.

The presence of additional constraints has driven the research field towards more specialized algorithms.
For example, in the streaming  scenario it is unfeasible to use the classical partitioning algorithm, since the data is continuously arriving.
A greedy algorithm that assign each incoming vertex to a partition has been proposed~\cite{Tsourakakis2012}
and computes partitions of only slightly less quality than most centralized algorithms.

Similarly, in the last decade the research field started investigating graph partitioning in distributed and 
decentralized systems. A few 
algorithms on distributed graph clustering have been developed, but they cannot be used for graph 
partitioning, since they do not obtain balanced partitions. Two candidate solutions are DIDIC and CDC:
DIDIC~\cite{Gehweiler2010} uses a diffusion process to move information across the graph and make sure that 
clusters are properly recognized, while CDC~\cite{Ramaswamy2005} simulates a flow of movement across
the graph to compute the community around an originator vertex. 

The algorithm selected for our comparison is JaBeJa~\cite{Rahimian2013},
a completely decentralized partitioning algorithm based on local and global
exchanges. Each vertex in the graph is initially mapped to a random partition. At each iteration, it will
try to exchange its mapping with one of its neighbor or with one of the random vertices obtained via a 
peer selection algorithm, if the exchange decreases the vertex cut size. An
additional layer of simulated annealing decrease the likelihood of returning to a
local minima. JaBeJa is similar in approach to Kernighan and Lin's algorithm,
but moves the choices from the partition level to the vertex level, greatly
increasing the possibility for parallelization.

\section{Conclusions}
\label{sec:conclusion}
This paper introduced the concept of edge partitioning for distributed graph analysis. Since solving
this problem exactly is unfeasible, we presented \dfep, an heuristic distributed edge partitioning algorithm
based on a simple funding model. The edge-partitioned graph resulting from \dfep can then be processed by \etsch, 
our graph processing framework.  Our experimental results, obtained through simulation and through an actual
deployment on an Amazon EC2 cluster, show that \dfep scales well and is able to obtain reasonably 
balanced partitions. Our implementation of \etsch in the Hadoop framework is much more efficient than the 
baseline solution, showing the promise of our approach.

As future work, we plan to thoroughly study the \etsch framework, both from a theoretical and a practical point 
of view. We plan to investigate how flexible the model is, to understand which type of graph problems are 
solvable and which ones need a completely different framework. For some problems, 
the classical solutions could be easily translated into \etsch, while for others novel algorithms could be needed. 
On the technical side, our Hadoop 
implementation of \etsch is still just a proof of concept. We plan to implement it more efficiently and study if 
other frameworks such as GraphLab, Stratosphere or Giraph lend themselves better to be the building block for 
\etsch.

\bibliographystyle{abbrv}
\bibliography{main}
\end{document}